\documentclass{article}

\usepackage{epsfig,amssymb,amsmath,graphicx,lineno}
\usepackage[compress]{cite}
\def\gsim{\mathrel\, {\vcenter {\baselineskip 0pt \kern 0pt\hbox{$>$} \kern 0pt \hbox{$\sim$} }}\,}

\author{Diego Harari, Silvia Mollerach and Esteban Roulet\\
{\it CONICET, Centro At\'omico Bariloche,}\\
{\it Avenida Bustillo 9500 (8400) Argentina.}}

\title{The shape of the extragalactic cosmic ray spectrum\\ from Galaxy Clusters}

\begin{document}
\maketitle

\begin{abstract}
  We study the diffusive escape of cosmic rays from a central source inside a galaxy cluster to obtain the suppression in the outgoing flux appearing when the confinement times get comparable or larger than the age of the sources. We also discuss the attenuation of the flux due to the interactions of the cosmic rays with the cluster medium, which can be sizeable for heavy nuclei.
  The overall suppression in the total cosmic ray flux expected on Earth is important to understand the shape of the extragalactic contribution to the cosmic ray spectrum for $E/Z<1$~EeV. This suppression can also be relevant to interpret the results of fits to composition-sensitive observables measured at ultra-high energies.

\end{abstract}

\section{Introduction}

Galaxy clusters are the largest virialized structures, having typical radii $R_{\rm cl}=2$--3~Mpc and masses  $M_{\rm cl}\sim 10^{13}$--$10^{15}M_\odot$. Most of the mass ($\sim 80$\%) consists of  Dark Matter while the baryonic mass is mostly ($\sim 80$\%) in diffuse gas which is quite hot ($T\sim 10^8$K) and dilute ($n_{\rm g}\simeq 10^{-4}$atoms~cm$^{-3}$), being observed by its X-ray emission from thermal bremsstrahlung. Clusters contain a large number of galaxies, $N_{\rm G}\simeq 10$--1000, with rich clusters having $N_{\rm G}>100$. They often have  active galaxies near their central parts and $\sim 20$\% of them have in their cores cD galaxies, which are giant ellipticals that can extend up to $\sim 300$~kpc. Clusters are already formed at redshifts $z\simeq 1$--2, and hence their ages  $t_{\rm cl}$ are comparable to the Hubble time $t_{\rm H}\equiv H_0^{-1}\simeq 14$~Gyr.  

Strong turbulent magnetic fields, with root mean square values $B\simeq {\rm few}\,\mu$G, are present inside clusters, having typical coherence lengths of 10--30~kpc. This implies that cosmic rays (CRs) accelerated in sources inside the clusters (such as supernovae or gamma ray bursts in starforming galaxies,  accretion shocks, jets or radiolobes in active galactic nuclei (AGN), etc.) can be confined for long times within the clusters. The confinement times are $t_{\rm esc}\equiv R_{\rm cl}^2/4D\simeq 30 (R_{\rm cl}/2\ {\rm Mpc})^2(10^{31}\ {\rm cm^2s^{-1}}/D)$~Gyr. Here the diffusion coefficient $D$ was scaled to its typical value inside a cluster for energies of 10~PeV. Note that $D$ increases with energy but $t_{\rm esc}$  may still exceed the age of the clusters 
even for energies as high as $ 10^{17}$--$10^{18}$~eV for protons, and $Z$ times larger for CRs with charge $Z$. This fact has led to consider clusters as reservoirs of CRs and considerable attention has  been devoted to the possibility that the CRs trapped in clusters could interact with the intracluster gas or radiation to produce secondary fluxes of $\gamma$ rays and neutrinos (see e.g. \cite{be97,ma06,ko09,mu13a,mu13,za15}). Although $\gamma$ rays from far away clusters would cascade down to energies of GeV--TeV by their interactions with the cosmic microwave background (CMB) and the infrared  and optical extragalactic background light (EBL), the neutrinos, which have typical energies $E_p/20$, with $E_p$ being the energies of the nucleons which produced them, could have a spectrum extending beyond the PeV range.  This could be interesting in view  of the recent results of the IceCube experiment that found an astrophysical contribution to the neutrino fluxes in the 20~TeV--few~PeV range \cite{ic}, although several constraints exist and the expected fluxes from interactions in the cluster medium are in general below the observed ones \cite{mu13,za15}. 
Note that the probability for the CR protons to interact with the diffuse cluster gas after propagating a time $t$ is usually modest, $P\simeq \sigma_{\rm pp}\overline{n}_{\rm g}ct\simeq 0.08(\overline{n}_{\rm g}/10^{-4} \mbox{cm}^{-3})(t/t_{H})$, where the proton-proton inelastic cross section at PeV--EeV energies is $\sigma_{\rm pp}\simeq 60$~mb and $\overline{n}_{\rm g}$
is the average gas density traversed.
This implies that the neutrino flux from CR interactions in the cluster gas is expected to be modest, except if CRs were to spend a lot of time confined in regions quite close to the cluster center where the gas density can be 
 larger, ${n}_{\rm g}>10^{-3} \mbox{cm}^{-3}$.  
The production of secondaries in p$\gamma$ interactions has also been considered \cite{ma06,ko09}, accounting for the enhanced IR and optical backgrounds inside the cluster, but given the energies of these photons the resulting neutrino fluxes only contribute eventually at $E_\nu>10$~PeV.

In this work we want to study the impact that the CR confinement in clusters could have for the spectrum of the extragalactic component of the CRs observed at the Earth, in the plausible scenarios in which the majority of the CR sources are indeed contained inside clusters. 
The main expected signature for this would be a suppression in the CR flux that would appear below the energy for which the CRs start to become unable to get out from the clusters that contain the sources. This effect should be rigidity dependent, suppressing the spectra of different nuclear components below an energy proportional to their respective charges. Besides the above mentioned effect, the enhancement of the confinement times at low  energies also leads to a larger attenuation of the CR fluxes due to interactions with cluster gas and radiation, which can be relevant in particular for heavy nuclei.

Although it may be hard to observe separately the extragalactic component below EeV energies due to the  predominance of the galactic CRs at these energies, this suppression could actually be useful to account for certain properties of the CR composition that were inferred  by the Auger Collaboration at higher energies \cite{augercomposition1,augercomposition2}. Indeed, by studying the distribution of the maximum depth of the air showers $X_{\rm max}$ determined by the fluorescence detectors of the Auger Observatory, a transition  towards heavier elements was found for energies above $\sim 5$~EeV  (as  inferred from the fact that the increase in the average value of $X_{\rm max}$ with increasing energies is slower than what would be expected for a pure proton composition). Moreover, a significant suppression of the abundance of heavy elements at energies below $\sim Z$~EeV is also suggested to account for the moderate values observed for the dispersion of $X_{\rm max}$, $\sigma(X_{\rm max})$, which requires to avoid large admixtures between light and heavy elements at a given energy.

The increasingly heavier composition for increasing energies is naturally explained in scenarios in which the CR spectra have a rigidity dependent cutoff, with a maximum energy at the sources of order $5 Z$~EeV.
Possible proposed ideas that could account for a light composition at few EeV energies and also lead to a suppression in $\sigma(X_{\rm max})$ are for instance to assume that the source spectra are quite hard, ${\rm d}\Phi_{\rm CR}/{\rm d}E\propto E^{-\gamma}$ with $\gamma$ in the range 1-1.6 \cite{Allard2011,Taylor2014,Aloisio2014,diMatteo2015}, to invoke the effects of photodisintegrations of heavy CR nuclei at the source environments \cite{un15} or to suppress the spectrum at low rigidities by the effects of diffusion in the extragalactic magnetic fields \cite{mo13}. Here we explore in detail the possibility  that this suppression could alternatively (or additionally) be due to the confinement of the low energy CRs directly in the clusters containing the sources. In all these scenarios the resulting  relative  suppression of the heavier components at the energies for which the light component dominates is crucial to reduce the spread in the values of $X_{\rm max}$ observed and should also help to reproduce the observed overall spectrum.

\section{Cosmic ray escape from galactic clusters}

The amount of time that CRs remain confined inside a galactic cluster depends on the properties of the diffusion through the turbulent magnetic field present in the cluster. 
The way the magnetic field power is distributed over different length scales, or equivalently wavenumbers $k$ (with $|B(k)|^2\propto k^{2-\alpha}$) is not known, but common assumptions are that it follows a Kolmogorov spectrum,  with $\alpha=1/3$, or a Kraichnan one,  with $\alpha=1/2$. The cluster magnetic fields have typical root mean square values $B\simeq 1$--10~$\mu$G and coherence lengths $l_{\rm c}\simeq 10$--30~kpc, where the coherence length is a fraction of the maximum scale of the turbulence $L_{\rm max}$, with e.g. $l_{\rm c}\simeq L_{\rm max}/5$ for Kolmogorov turbulence \cite{ac99,ha02}.

The critical energy $E_{\rm c}=ZeBl_{\rm c}\simeq 9Z(B/\mu\mbox{G})(l_{\rm c}/10\,\mbox{kpc})$~EeV is the energy at which the effective Larmor radius  $r_L\equiv E/ZeB $ equals the coherence length. This critical energy then separates the low energy regime of resonant diffusion, in which particles can be strongly deflected by scattering from $B$ field modes at the scale of their Larmor radius, from the high energy regime in which  the deflections in each coherence length are small and the scattering is non-resonant.
We see that inside a cluster the CRs suffer resonant diffusion up to quite high energies, which even in the case of protons is larger than the energy of the ankle of the CR spectrum, which is where the hardening observed above $\sim 5$~EeV begins. 

The associated diffusion coefficient $D$ can be obtained from fits to the results of numerical simulations of charged particle trajectories in turbulent magnetic fields, leading to \cite{I}
\begin{equation}
D=\frac{c}{3}l_{\rm c}\left[4\left(\frac{E}{E_{\rm c}}\right)^2+a_I\left(\frac{E}{E_{\rm c}}\right)+a_L\left(\frac{E}{E_{\rm c}}\right)^{\alpha}\right],
\end{equation}
where for a Kolmogorov spectrum one finds that $a_I=0.9$ and $a_L=0.23$ while for the Kraichnan case  $a_I=0.65$ and $a_L=0.42$.
Focusing for definiteness on the case of a Kolmogorov spectrum one finds  that for $E\ll E_{\rm c}$, which is the regime we will be interested in, 
\begin{equation}
D\simeq a_L \frac{cl_{\rm c}}{3}\left( \frac{E}{E_{\rm c}}\right)^\alpha\simeq 3.4\times 10^{31} \left(\frac{l_{\rm c}}{10\,\mbox{kpc}} \right)^{2/3} \left(\frac{\mu\mbox{G}}{B} \right)^{1/3} \left(\frac{E/Z}{\rm EeV} \right)^{1/3} \ {\rm \frac{cm^2}{s}}.
\label{eq:dturb}
\end{equation}
One should keep in mind however that for $E/E_{\rm c}>0.05$ it is convenient to use the complete energy dependence of the diffusion coefficient.

We will consider for simplicity  the case of a source located at the center of a spherical cluster and adopt an homogeneous turbulent field, so that $D$ is spatially constant. This will enable us to obtain analytical expressions which are useful to understand the main relevant issues. We will later comment on the  impact of having a radially varying
magnetic field, in which case it is also helpful to obtain numerically the CR trajectories to understand the resulting behavior.

 The diffusion equation describing the evolution of the CR density for spatially constant $D$ is
\begin{equation}
\frac{\partial n}{\partial t}=\frac{D}{r^2}\frac{\partial}{\partial r}\left( r^2\frac{\partial n}{\partial r}\right)+\frac{\partial}{\partial E}(b\,n)+Q,
\end{equation}
where $b=-{\rm d}E/{\rm d}t$ accounts for energy losses. We will first consider the case in which energy losses are negligible, which is a reasonable approximation for protons and light nuclei in the energy range in which we are interested,  $E/Z<1$~EeV. We will discuss in Section~3  the additional attenuation effects induced by the interactions with the cluster gas and radiation, which can be sizeable for heavy nuclei and can further suppress the escape of the CRs from the cluster. 
Note also that since clusters are already gravitationally bound since very early times, and hence are decoupled from the Universe's expansion, the CR energy will not be redshifted during the diffusion process inside the cluster. Only later when the CRs travel from the cluster to the Earth do their redshift effects need to be taken into account.

If CRs can freely escape outwards once they reach the border of the cluster, the boundary condition will be\footnote{Actually the density doesn't exactly vanish at the cluster radius, since indeed a non-zero CR flux is present there, but the density is negligible with respect to the one that would be obtained at this radius in the case of an unbounded diffusing medium.} that $n(R_{\rm cl},t)=0$. The Green's function $G$ for the diffusion equation, i.e. the solution,  valid for $r<R_{\rm cl}$, for a source term $Q=\delta(\vec{r})\delta(t)$,  is \cite{greenf}
\begin{equation}
G_{<}(r,t)=\frac{1}{R_{\rm cl}^3(\pi \tau)^{3/2}}\sum_{n=-\infty}^\infty \frac{x_n}{x}\exp[- x_n^2/\tau],
\label{eq:gmen}
\end{equation}
where $x_n\equiv x+2n$ with $x\equiv r/R_{\rm cl}$ being the rescaled distance from the central source and $\tau\equiv t/t_{\rm esc}$, with $t_{\rm esc}\equiv R_{\rm cl}^2/4D$. Note that there is an energy dependence which arises through the diffusion coefficient itself and that the $n=0$ term in the sum just gives the known contribution for the case of diffusion in an infinite medium. The previous expression efficiently converges for $t<t_{\rm esc}$ while the alternative expression
\begin{equation}
G_{>}(r,t)=\frac{1}{2xR_{\rm cl}^3}\sum_{m=1}^\infty m\sin(m\pi x)\exp\left[ -\left(\frac{ m\pi}{2}\right)^2\tau\right]
\label{eq:gmay}
\end{equation}
has a better convergence for $t>t_{\rm esc}$.
If we consider now a source which turned on at $t=0$, emitting $Q$ CRs per unit time, the density of CRs can be obtained by integrating in time the Green's functions weighted by the source term, i.e.
\begin{equation}
n(r,t)=\int_0^{t}{\rm d}t'\, Q(t') G(r,t-t').
\end{equation}
Considering a steady source, i.e. with constant $Q$ for $t>0$, 
 one then obtains for $t<t_{\rm esc}$
\begin{equation}
n_{<}(r,t)=\frac{Qt_{\rm esc}}{\pi R_{\rm cl}^3 x}\sum_{n=-\infty}^\infty s_n {\rm Erfc}\left(\frac{|x_n|}{\sqrt{\tau}}\right),
\end{equation}
where $s_n\equiv x_n/|x_n|$ is the sign of $x_n$ and 
 $\mbox{Erfc}(x)=2\int_x^\infty {\rm d}y\exp(-y^2)/\sqrt{\pi}$ is the complementary error function.

On the other hand, for $t>t_{\rm esc}$ it is convenient to use the expression
\begin{equation}
n_{>}(r,t)=n_{<}(r,t_{\rm esc})+\frac{Qt_{\rm esc}}{\pi R_{\rm cl}^3 }\frac{2}{\pi x}\sum_{m=1}^\infty \frac{\sin(m\pi x)}{m}\left(\exp\left[-\left(\frac{m\pi}{2}\right)^2\right]-\exp\left[-\left(\frac{m\pi}{2}\right)^2\tau\right]\right).
\end{equation}

\begin{figure}[ht]
\centerline{\epsfig{width=3.in,angle=0,file=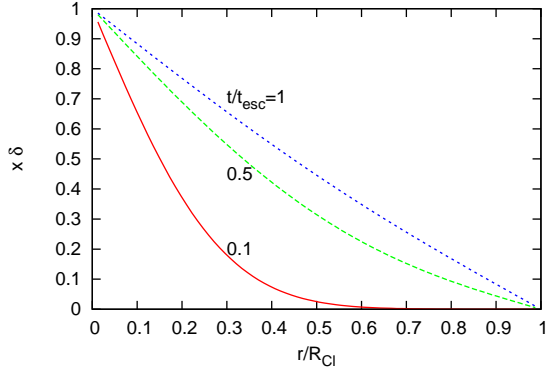}}
\caption{Normalized density $\delta$ multiplied by $x$ as a function of $x\equiv r/R_{\rm cl}$ for different values of $t/t_{\rm esc}$, as indicated.} 
\label{fig:dens}
\end{figure}

In Fig.~\ref{fig:dens} we show the quantity $x\delta$, where the normalized density is $\delta(r,t)\equiv n(r,t)/(Q t_{\rm esc}/\pi R_{\rm cl}^3)$, as a function of $x\equiv r/R_{\rm cl}$ and for different values of $\tau\equiv t/t_{\rm esc}$. For small $\tau$ the density is quite centrally concentrated but then progressively converges to the asymptotic value corresponding to $t\gg t_{\rm esc}$, which is $\delta(\tau\gg 1)\simeq (1-x)/x$. 

From this density one may now obtain the CR current throughout the cluster as $\vec{J}=-D\vec{\nabla} n$, which points radially outwards. By direct computation we find its amplitude to be
\begin{equation}
J_{<}(r,t)=\frac{Q}{4\pi r^2}\sum_{n=-\infty}^\infty \left\{s_n {\rm Erfc}\left(\frac{|x_n|}{\sqrt{\tau}}\right)+\frac{2x}{\sqrt{\pi \tau}}\exp\left[-\frac{x_n^2}{\tau}\right] \right\}
\end{equation}
for $t<t_{\rm esc}$, while
\begin{equation}
J_{>}(r,t)=J_{<}(r,t_{\rm esc})+\frac{Q}{4\pi r^2 }\frac{2}{\pi}\sum_{m=1}^\infty \frac{\sin(m\pi x)-m\pi x\cos(m\pi x)}{m}\left(\exp\left[-\left(\frac{m\pi}{2}\right)^2\right]-\exp\left[-\left(\frac{m\pi}{2}\right)^2\tau\right]\right)
\end{equation}
for $t>t_{\rm esc}$.

We may now introduce the suppression factor
\begin{equation}
\eta\equiv \frac{J(R_{\rm cl},t)}{Q/(4\pi R_{\rm cl}^2)},
\end{equation}
which determines the amount by which the source luminosity is modified by the confinement inside the cluster.

\begin{figure}[ht]
\centerline{\epsfig{width=3.in,angle=0,file=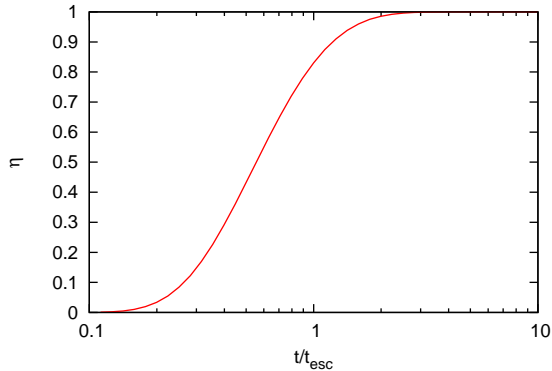}}
\caption{Suppression factor $\eta$  as a function of $t/t_{\rm esc}$.} 
\label{fig:etavst}
\end{figure}

In  Fig.~\ref{fig:etavst} we show this  factor as a function of $t/t_{\rm esc}$.
The main suppression effect is then due to the fact that at energies for which $t_{\rm esc}\geq t$ the fraction of CRs which have a non-negligible chance of escaping becomes very small (and only those emitted at very early times can eventually do so). Note that both Fig.~\ref{fig:dens} and \ref{fig:etavst} are independent from the turbulence spectrum, which only enters in the relation between the escape time and the energy.

\begin{figure}[ht]
\centerline{\epsfig{width=3.in,angle=0,file=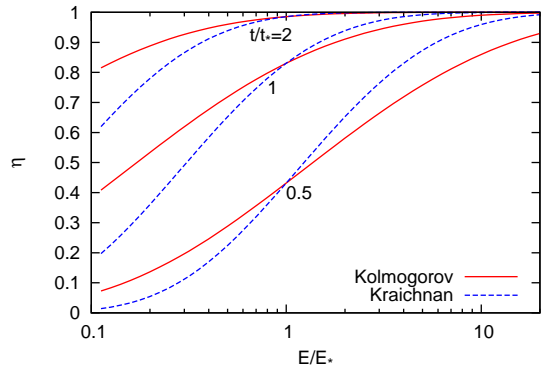}}
\caption{Suppression factor $\eta$  as a function of $E/E_*$ for different values of $t/t_*$, with $t_*=t_{\rm esc}(E_*)$.} 
\label{fig:etavse}
\end{figure}
In  Fig.~\ref{fig:etavse} we show the suppression factor as a function of the energy, where we set $t_{\rm esc}\equiv t_*(E/E_*)^{-\alpha}$, with $E_*$ just an arbitrary  reference energy and $t_*$ being the value of $t_{\rm esc}$ at this reference energy. Solid lines are for the Kolmogorov case ($\alpha=1/3$) while dashed lines are for the Kraichnan turbulence ($\alpha=1/2$). The different curves in each case correspond to values of $t/t_*$ of  0.5,  1 and 2, and it is apparent that the suppression becomes stronger for shorter times. One can also see that for a given $E_*$, if $t/t_*$ is scaled by a factor $\kappa$ the corresponding curve is displaced in energy by a factor $\kappa^{1/\alpha}$.

It is convenient to define,  for a given $t$, the escape energy $E_{esc}$ as the one below which the suppression becomes strong, which is obtained 
from the condition $t_{\rm esc}(E_{\rm esc})=t$.
For instance, in the case of a Kolmogorov spectrum one has that
\begin{equation}
t_{\rm esc}=\frac{R_{\rm cl}^2}{4D}\simeq 9\,{\rm Gyr}\, \left(\frac{E/Z}{\rm EeV} \right)^{-1/3} \left(\frac{R_{\rm cl}}{2\,\rm Mpc} \right)^2 \left(\frac{B}{\mu{\rm G}} \right)^{1/3} \left(\frac{l_{\rm c}}{10\,{\rm kpc}} \right)^{-2/3},
\label{eq:tesc}
\end{equation}
so that
\begin{equation}
E_{\rm esc}\simeq Z\left(\frac{9\,{\rm Gyr}}{t}\right)^3\,  \left(\frac{R_{\rm cl}}{2\,\rm Mpc} \right)^6 \left(\frac{B}{\mu{\rm G}} \right) \left(\frac{l_{\rm c}}{10\,{\rm kpc}} \right)^{-2} \,{\rm EeV}.
\end{equation}
Analogous expressions can be directly obtained for other turbulence spectra.

Note that the escape energy has a strong dependence on the cluster radius, and actually the relevant size is the one defining the region where the magnetic field is sizeable so that the confinement is effective. The assumption we made of taking  $B$ as constant within the typical cluster virial radius  is certainly an idealized one.  In principle, the explicit radial dependence of the magnetic field strength may be taken into account.
However, the precise cluster magnetic field profiles are largely unknown. For instance, cosmological simulations of a compact  intermediate mass cluster with primordial seed magnetic fields \cite{du08} obtained that the magnetic field resulting from the collapse could be quite suppressed in the outskirts of the cluster,  with $B\ll\mu$G beyond 0.5~Mpc from the  center. Adopting such models, as in \cite{ko09},  should lead to escape energies not much larger than $Z$~PeV for $t\sim t_{\rm H}$.  On the other hand, actual fits to observed Faraday rotation measures in the Coma Cluster \cite{bo10} imply that the strength of its magnetic field, which is  $\sim 5\,\mu$G at the core, remains larger than $0.5\,\mu$G up to at least 2~Mpc radius. This indicates that for cosmological times
one could expect indeed that  $E_{\rm esc}\sim(0.1$--$1)Z$~EeV, as suggested by the simplified models we considered before.

Note that in the case of a constant $B$ field
the escape time, defined as $t_{\rm esc}=R_{\rm cl}^2/4D$, is the one for which a fraction $f=83$\% of the emitted CRs in a burst at $t=0$ would have already exited the cluster. This can be seen by explicit integration of the Green's function in eq.~(\ref{eq:gmen})
over the cluster volume, since one should have
\begin{equation}
f=1-\int_0^{R_{\rm cl}}{\rm d}r\,4\pi r^2G(r,t).
  \end{equation}
We may then adopt the time at which 83\% of the CRs from a bursting source exit the cluster as the definition of the escape time also in the case in which the magnetic field, and hence the diffusion coefficient, has a radial dependence.
Equivalently, $t_{\rm esc}$ is the time for which $\eta=0.83$, i.e. at which the flux
integrated over  the cluster boundary which is due to a central steady source that started at $t=0$ equals 83\% of the source luminosity. This can be directly proved by noting that
\begin{equation}
 4\pi R^2 J(R,t)=Q\int_0^t{\rm d}t'\,\left[-\frac{\rm d}{{\rm d}t'}\int_0^R{\rm d}r\,4\pi r^2G(r,t-t')\right],
\end{equation}
and hence
\begin{equation}
  \eta(t)=\frac{J(R_{\rm cl},t)}{Q/4\pi R_{\rm cl}^2}=1-\int_0^{R_{\rm cl}}{\rm d}r\,4\pi r^2G(r,t)=f.
  \end{equation}

Actually, we find that an accurate  expression for the escape time is
\begin{equation}
  t_{esc}\simeq\frac{1}{2}\int_0^{R_{\bf cl}} {\rm d}r\,\frac{r}{D(r)}.
  \label{eq:tescapp}
\end{equation}
To derive it, let's consider first the asymptotic steady regime at large times for which the outward CR flux through the cluster boundary equals the production rate inside it.  One  has in this case that, for arbitrary radial dependence of the magnetic field, 
\begin{equation}
  J=-D(r)\frac{\partial n}{\partial r}=\frac{Q}{4\pi r^2},
\end{equation}
and hence 
\begin{equation}
 \frac{\partial n}{\partial r}=-\frac{Q}{4\pi r^2 D(r)}.
\end{equation}
Using that $n(R_{\rm cl})=0$ this can be directly integrated to obtain
\begin{equation}
  n(r)=\int_r^{R_{\rm cl}}{\rm d}r\,\frac{Q}{4\pi r^2 D(r)}.
\end{equation}
One may now exploit the fact that  the CR density at a given radius is just proportional to the average time spent by the diffusing CRs in a given volume element at that location. If we assume then that $t_{\rm esc}$ is proportional to the average time spent by the CRs inside the cluster, we should have
\begin{equation}
  t_{\rm esc}={\cal N}\int_0^{R_{\rm cl}}{\rm d}r\,4\pi r^2 n(r)=-\frac{4\pi}{3}{\cal N}\int_0^{R_{\rm cl}}{\rm d}r\,r^3 \frac{\partial n}{\partial r}=\frac{{\cal N}Q}{3}\int_0^{R_{\rm cl}}{\rm d}r\,\frac{r}{D},
\end{equation}
where we integrated by parts. The normalization factor ${\cal N}=3/2Q$ is chosen so as to reproduce the definition of $t_{\rm esc}$ for constant $D$.
Note that the average time spent by the diffusing CRs between $r$ and $r+\Delta r$ will be $\Delta t=r/(2 D(r))\Delta r$. 
Noting finally that the distribution of these times for the particles escaping before $t_{\mathrm esc}$ from a bursting source is already quite close to the expression just derived for the  asymptotic regime,  one obtains that the approximate expression in eq.~(\ref{eq:tescapp}) should hold.

 We can consider now the  case of the Coma Cluster  magnetic field that was  obtained in \cite{bo10}, which is
\begin{equation}
B=B_0 \left[ 1+\left(\frac{r}{r_{\rm c}}\right)^2\right]^{-0.56},
\label{eq:tescnum}
\end{equation}
with $B_0=4.7\,\mu$G and $r_{\rm c}=291$~kpc.
In this case and setting $B=0$ for $r>2$~Mpc, we obtain that  the expression in eq.~(\ref{eq:tescapp}) for $t_{\rm esc}$ gives a value 0.98 times the value corresponding to $B=1\,\mu$G within 2~Mpc, while if we were e.g. to assume that the boundary of the Coma Cluster is at 3~Mpc we would obtain an escape time about twice as large\footnote{Including the full energy dependence of $D$ according to eq.~(1), the escape time for the example of the Coma cluster up to 3~Mpc would be a factor 1.56 or 1.88 larger, for energies of 1~EeV or 0.01~EeV respectively, than the corresponding value for $B=1\,\mu$G  up to 2~Mpc.}.
We see then that the estimates obtained with constant magnetic fields up to the virial radius are indeed quite  reasonable and we will hence continue to use them in the following.

To further check these results we also performed simulations of  trajectories of CRs that diffuse from a central source by integrating a  stochastic differential equation  as in \cite{ac99,I} in this radially varying magnetic field, obtaining very good agreement with the analytical approximate results and hence validating the expression for $t_{\rm esc}$ in eq.~(\ref{eq:tescapp}).

\section{Effects of CR interactions inside the cluster}

Up to now we neglected the possible effects of the interactions of the CRs with the gas and radiation inside the cluster. Since at the energies under consideration the inelasticities of these interactions are large, we can adopt the simplified picture that every CR that follows an inelastic interaction is essentially lost (i.e. we don't follow in detail the secondary particles). In this case the interactions will have the effect of just further suppressing the outgoing CR flux by an amount $P\simeq\exp[-ct_a/\lambda]$, where $t_a={\rm min}(t,t_{\rm esc})$, with $t$ the time being considered and $\lambda$ being the appropriate attenuation length.

  Let us discuss first the case of protons, remembering that we are focusing on energies in the range 10~PeV--few~EeV. The interaction with the cluster gas, assumed to be mostly hydrogen, has an inelastic scattering cross section given by
  \begin{equation}
\sigma_{\rm pp}\simeq [32.4-1.2{\rm ln}s+0.21({\rm ln}s)^2]\,{\rm mb},
  \end{equation}
  where the squared center of mass energy $s\simeq 2m_pE$ is expressed in GeV$^2$. The attenuation length for this interaction will be $\lambda_{\rm pp}=(\sigma_{\rm pp}\overline{n}_{\rm g})^{-1}$, with $\overline{n}_{\rm g}$ being the average gas density traversed by the CRs in their path from the central source until they exit the cluster. 
  The gas density in a cluster is usually parameterized by a beta model, according to
  \begin{equation}
    n_{\rm g}(r)=n_0 \left(1+\frac{r^2}{r_{\rm c}^2}\right)^{-3\beta/2}.
  \end{equation}
  For instance, for the Coma cluster one can take $n_0=3.4\times 10^{-3}$~cm$^{-3}$, $r_{\rm c}=291$~kpc and $\beta=3/4$ \cite{bo10}. 
The average gas density can be estimated, following a similar approach as in Section~2, as
\begin{equation}
  \overline{n}_{\rm g}\simeq \frac{1}{t_{\rm esc}}\int_0^{R_{\rm cl}}{\rm d}r\ \frac{r}{2D(r)}n_{\rm g}(r),
\end{equation}
  leading to  $\overline{n}_{\rm g}\simeq 3\times 10^{-4}$cm$^{-3}$ using the magnetic field and gas density of the Coma cluster.  Note that if we were to use this gas density but a constant magnetic field of $1\,\mu$G, we would get  $\overline{n}_{\rm g}\simeq 2.2\times 10^{-4}$cm$^{-3}$ since the relative times spent at different radii would be different than for the case of a radially dependent $B$ field.

\begin{figure}[ht]
\centerline{\epsfig{width=2in,angle=0,file=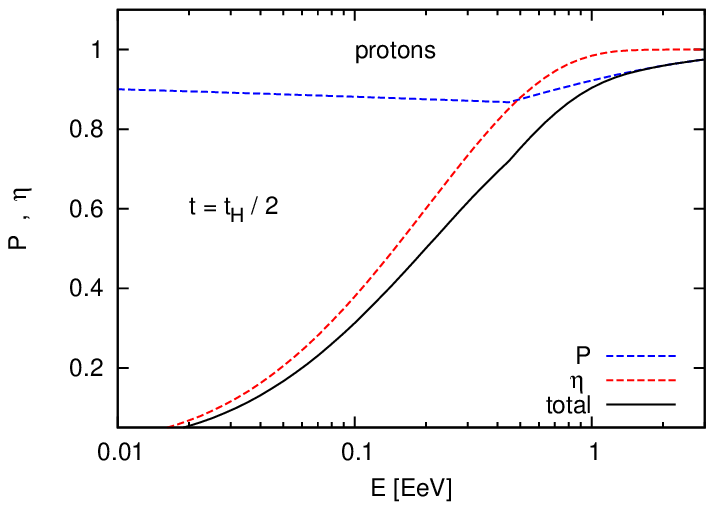}\epsfig{width=2in,angle=0,file=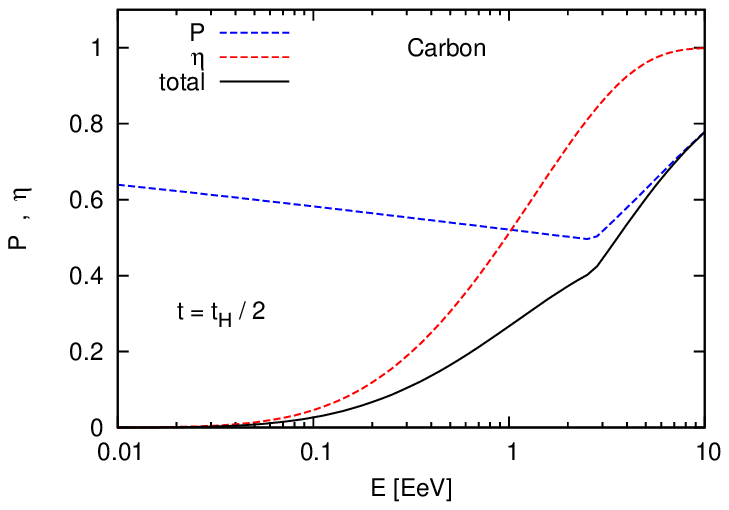}\epsfig{width=2in,angle=0,file=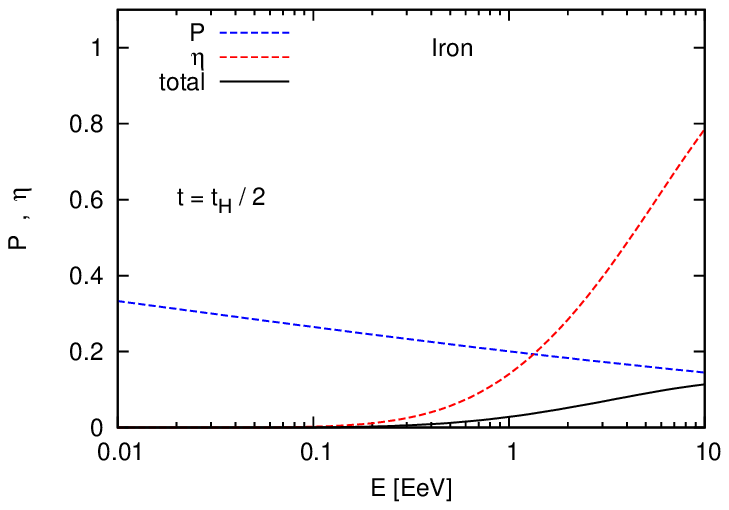}}
\caption{Suppression due to the magnetic confinement ($\eta$) and to the hadronic interactions with the gas ($P$), as well as the total one. Left panel is for protons, medium one for C  and right panel for Fe nuclei. They are shown  as a function of $E$ adopting $t=t_{\rm H}/2$.} 
\label{fig:attnucvse}
\end{figure}

 We show in the left panel of Fig.~\ref{fig:attnucvse}  the resulting suppression as a function of the energy  adopting a constant value $B=1\,\mu$G inside 2~Mpc and for  $t=t_{\rm H}/2$ (we use here the full energy dependence of the diffusion coefficient and of the inelastic pp cross section). For comparison, also the suppression $\eta$ of the flux due to the magnetic confinement discussed before is shown as well as the total suppression, which is given to a good approximation by the product of the confinement and interaction suppressions.
 Note that had we considered smaller times $t$ the confinement suppression would extend up to higher energies, as was illustrated in Fig.~\ref{fig:etavse}, and the interaction suppression would have been even less relevant since for  $t<t_{\rm esc}$ the CRs that manage to escape within $t$  would have traversed a smaller column density of target material.

 Besides the hadronic interactions with the gas, protons may interact with background photons. Photopion production interactions with the CMB are only relevant for energies larger than 30~EeV, and hence can be ignored, while the pair creation losses have quite large attenuation lengths. For proton energies above an EeV there could also be the photopion production off IR and optical photons produced by the stars in the cluster itself, that may lead to an average  photon density inside the cluster larger than the EBL by about an order of magnitude or so \cite{ma06,ko09}. If one ignores possible spectral differences between the photon background of the cluster (which is mostly due to the emission by stars in elliptical galaxies) and that of the EBL (which also gets a contribution from IR emission from the gas in spiral galaxies), one could model this attenuation length just by scaling the attenuation corresponding to the present day EBL. In particular, when  considering  an average photon density 10 times higher than the EBL, i.e. scaling the EBL attenuation length by a factor of 0.1, this would  anyway lead to an attenuation length much larger than that caused by the hadronic interactions that we already considered and hence this contribution can  also be ignored.

 Turning now to the case of nuclei, here the effects of interactions are larger. The inelastic hadronic interactions indeed scale approximately as $\sigma_{\rm pA}(E)\simeq A^{2/3}\sigma_{\rm pp}(E/A)$, and hence the attenuation is stronger if the mass number $A$ is large. We show in the middle and right panels of Fig.~\ref{fig:attnucvse} the resulting attenuation factors for the cases of C and Fe respectively, for $t=t_{\rm H}/2$ and the same cluster parameters as in the proton example. Not only is the cross section of nuclei larger than that for protons, but also the escape times at a given energy become larger for nuclei than for protons due to the larger electric charges $Ze$. The confinement effect will dominate at low energies, i.e. when $t_{\rm esc}> t$, while the attenuation effect due to the interactions can become dominant at the high energies where the escape times become small compared to the time considered. 

\begin{figure}[ht]
\centerline{\epsfig{width=3.in,angle=0,file=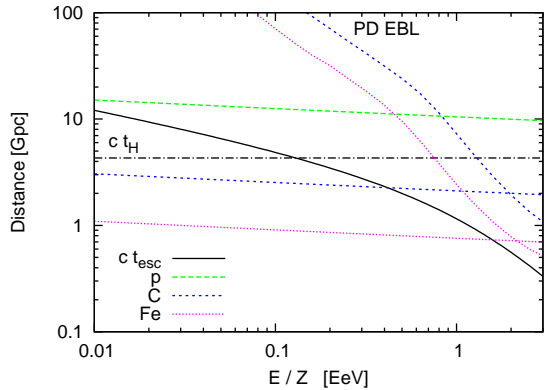}}
\caption{Attenuation lengths for hadronic interactions of protons, C and Fe   as a function of $E/Z$. Also shown is the photodisintegration energy loss length for C and Fe off the EBL (PD EBL lines). Solid line is the escape length $ct_{\rm esc}$ and horizontal dot-dashed line is the horizon distance. } 
\label{fig:tescvse}
\end{figure}

Regarding the interactions with the photon background, the dominant attenuation in this energy range will be the nuclear photodisintegration off the IR and optical photons. We 
show in Fig.~\ref{fig:tescvse} the different attenuation lengths for protons, C and Fe due to the hadronic interactions, as well as the escape length $ct_{\rm esc}$ and the horizon distance $ct_{\rm H}$. For the nuclei we also show the energy loss lengths for interactions with the EBL as computed in \cite{ha15}. Note that  if one considers that the cluster background radiation traversed by the CRs has an average density ten times larger than the EBL, the resulting attenuation length would be 10 times smaller than those shown in the figure for the case of the EBL. This would make the attenuation of the nuclei due to photodisintegrations more relevant than the ones due to hadronic interactions at energies $E/Z>0.5$~EeV, and since we are focusing here on rigidities lower than this we can hence disregard those interactions.

\section{CR spectrum from all the clusters}

To obtain the extragalactic contribution to the CR spectrum observed at Earth
one needs to add the contributions from all the sources at different distances, each one suppressed by the corresponding factors associated to the times at which the observed CRs left the sources. One can also include redshift effects, but we will neglect however energy losses due to interactions during the propagation from the sources, which is reasonable for $E/Z<{\rm EeV}$,  and also assume that  the possible additional effects of diffusion in extragalactic magnetic fields are subdominant.
If we consider for simplicity that the clusters are homogeneously distributed in comoving space, that the sources inside each of them have the same steady luminosities $Q$,  and that they have been emitting continuously since a redshift $z_{\rm m}$ (typically $z_{\rm m}=1$--2), the CR flux at Earth can be obtained as 
\begin{equation}
  \frac{{\rm d}\Phi}{{\rm d}E}=\frac{c}{4\pi}\int_0^{z_{\rm m}}{\rm d}z\, \left|\frac{{\rm d}t}{{\rm d}z}
  \right|\eta(t,E_{\rm g})\,P(E_{\rm g})\,\frac{{\rm d}Q}{{\rm d}E_{\rm g}}\frac{{\rm d}E_{\rm g}}{{\rm d}E}n_{\rm cl},
  \label{eq:flux}
\end{equation}
where $E_{\rm g}=(1+z)E$ is the energy at the source of the CR which arrives with energy $E$. For the differential spectrum at the sources we adopt a power law d$Q/{\rm d}E_{\rm g}\propto E_{\rm g}^{-\gamma}$ and $n_{\rm cl}$ is the comoving cluster density. On the other hand
\begin{equation}
\frac{{\rm d}t}{{\rm d}z}=-t_{\rm H}\frac{1}{(1+z)\sqrt{(1+z)^3\Omega_{\rm m}+\Omega_\Lambda}},
\end{equation}
where $\Omega_{\rm m}=0.3$ and $\Omega_\Lambda=0.7$ are the adopted values of the present matter and dark energy contributions to the cosmological density. In eq.~(\ref{eq:flux}) the suppression of the cluster luminosity $\eta(t)$ is to be evaluated at a time
\begin{equation}
t=\int_z^{z_{\rm m}}{\rm d}z'\,\left|\frac{{\rm d}t}{{\rm d}z'}\right|.
\end{equation}
The factor $P(E_{\rm g})$ includes the further suppression due to the interactions with the cluster medium.

We can then obtain the overall suppression of the CR flux observed at the Earth induced by the different suppressions incurred in the individual clusters at the different redshifts as
\begin{equation}
S(E)\equiv \frac{{\rm d}\Phi/{\rm d}E}{({\rm d}\Phi/{\rm d}E)_{\eta P=1}},
\end{equation}
where the denominator is the flux that would be obtained for $\eta P=1$, i.e. in the absence of diffusion and interaction effects in the clusters.

\begin{figure}[ht]
\centerline{\epsfig{width=3in,angle=0,file=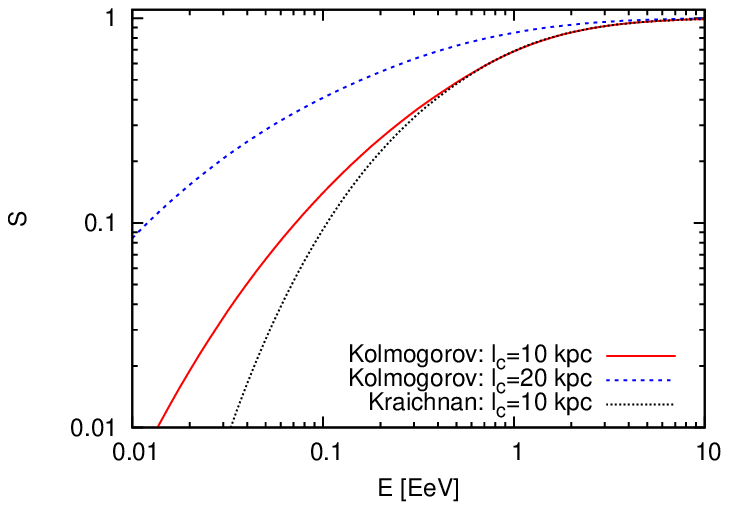}\epsfig{width=3in,angle=0,file=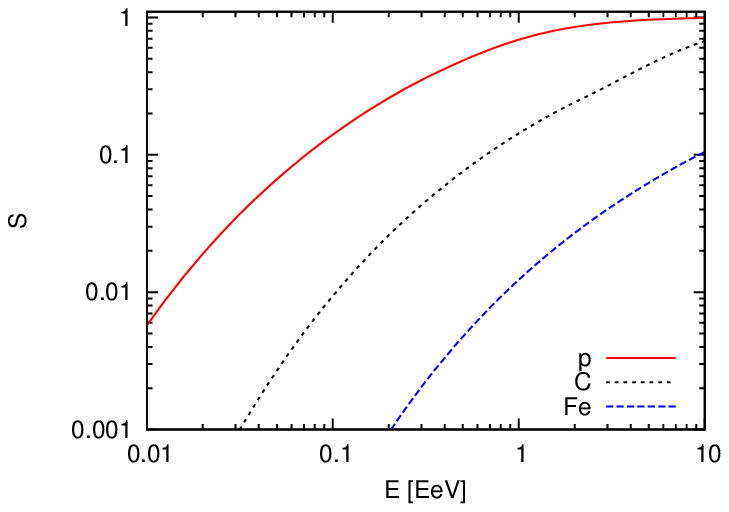}}
\caption{Suppression factor $S$  as a function of $E$. Left panel is for protons and different assumptions on the magnetic fields. Right panel is for p, C and Fe for Kolmogorov turbulence, $l_{\rm c}=10$~kpc and $B=1\,\mu$G.} 
\label{fig:attvse}
\end{figure}

The results of the evaluation of the suppression factor are shown as a function of the energy  in Fig.~\ref{fig:attvse}, adopting $B=1\,\mu$G, $R_{\rm cl}=2$~Mpc, $\gamma=2$ and $z_{\rm m}=1$ (being the results for $z_{\rm m}=2$  qualitatively very similar).  The left panel shows the case of protons for Kolmogorov turbulence with $l_{\rm c}=10$ and 20~kpc, as well as the case of Kraichnan turbulence with $l_{\rm c}=10$~kpc, which leads to a steeper shape for the suppression. The right panel in the figure shows, for Kolmogorov turbulence with $l_{\rm c}=10$~kpc, the results for protons, carbon and iron nuclei.
Note also that  the suppressions obtained  do not depend on the actual density of the CR sources or on their assumed luminosity $Q$.
A simple fit to these curves of the form $S=\exp(-a/E^b)$ leads to $a=0.35$, 2.0 and 4.4 while $b=0.83$, 0.6 and 0.3 for p, C and Fe respectively.
These expressions may prove helpful when performing fits to the observed composition data.

It is apparent from the figures that the suppression for the case of protons is already noticeable  for energies of few EeV and it becomes quite strong, $S<0.1$, for $E<0.1$~EeV. The slope of the suppression will have the effect of hardening the observed flux with respect to the source flux, with the change in the spectral slope depending on the energy considered. At $E\simeq 0.1$~EeV the effective spectral index (i.e. the logarithmic slope of the spectrum) should already become larger by an amount of order unity, and would be larger for the Kraichnan than for the Kolmogorov turbulence spectrum.

In the case of nuclei the suppression already appears at higher energies, since  the effect due to the confinement in the clusters is rigidity dependent. However, at $E>1$~EeV  the main attenuation of the fluxes of nuclei is due to the interactions with the cluster gas, and at even higher energies, $E>0.5Z$~EeV it would actually be the interactions with the optical and IR radiation in the cluster, not included here, the dominant ones. We see from  Fig.~\ref{fig:attvse} that the fluxes of heavy nuclei are quite depleted, with $S<0.1$ for $E/Z<0.3$~EeV, and the spectra of the CRs reaching Earth would look quite hard at these energies due to the cluster confinement and energy loss effects.

We note that the details of the suppression obtained depend on several assumptions, such that the clusters are homogeneously distributed and have similar characteristics (magnetic fields, cluster sizes, similar gas densities, central steady sources of common luminosities, maximum redshift, etc.). In particular, the precise value of the energy below which the suppression becomes sizeable  and also the actual shape of this attenuation will depend on the actual values of all these parameters.


\section{Discussion}

The suppression appearing at low rigidities in the expected CR spectrum due to the confinement in galaxy clusters, eventually enhanced by the attenuation due to the CR interactions with the gas or radiation present in it, can be relevant to understand some observations. In particular, at EeV energies the CRs appear to be light, with protons  likely contributing a dominant fraction \cite{augercomposition1,augercomposition2}. On the other hand, the small level of large scale anisotropies, with the dipolar component of the CR arrival direction distribution being not much larger than the percent level at EeV energies \cite{augerLS2013}, suggests that this light component is unlikely to be of galactic origin. An extragalactic light component may emerge, somewhere in the energy range from 0.1--1 EeV, above the heavy galactic component present at lower energies. On the other hand, composition  determinations by the Auger Observatory indicate that above $\sim 5$~EeV the CRs become increasingly heavier, suggesting the presence of a rigidity dependent cutoff in the source spectra at $E/Z\sim 5$~EeV. Since above 5~EeV and up to the high energy suppression starting above 30~EeV the total spectrum behaves as d$\Phi/{\rm d}E\propto E^{-2.6}$, this means that the emerging heavier components should have also contributed significantly at EeV energies, something which doesn't seem to be the case, unless their spectrum were very hard. Given all these constraints from the spectrum and the composition, it was found that the individual spectra at the sources of the different elements, parametrised as power laws d$\Phi/{\rm d}E\propto E^{-\gamma}$ below their cutoff values,  should be quite hard, with best fits for $\gamma$ in the range 1-1.6 \cite{Allard2011,Taylor2014,Aloisio2014,diMatteo2015}. These hard spectra would also help to account for the moderate values observed of $\sigma(X_{\rm max})$, since the admixture between components with very different masses gets reduced. Having such hard spectra at the sources is however not very natural because the usual Fermi acceleration leads to $\gamma\simeq 2$ and inefficiencies or energy losses can only steepen the spectrum. On the other hand, the low energy suppression discussed in this paper  could naturally explain these observations, with the apparent hardening of the observed spectra being just due to the fact that as the energy is lowered, the CRs have an increasing difficulty  to escape from the clusters that host their sources.

\section*{Acknowledgments}
Work supported by CONICET (grant PIP 2012-11220110100447) and ANPCyT (grant PICT 2013-0213), Argentina.
We thank the members of the Auger Collaboration for 
useful discussions.

\end{document}